\documentclass[prd,reprint,showpacs,showkeys]{revtex4-1}
\usepackage{amsfonts}
\usepackage{amssymb}
\usepackage{amsmath}
\usepackage{graphicx}
\usepackage{float}
\usepackage{subcaption}
\usepackage[font={footnotesize,it}]{caption}

\begin{document}

\title{Note on an elementary particle model with Bertotti-Robinson core}
\author{S. Habib Mazharimousavi}
\email{habib.mazhari@emu.edu.tr}
\author{M. Halilsoy}
\email{mustafa.halilsoy@emu.edu.tr}
\affiliation{Department of Physics, Eastern Mediterranean University, Gazima\u{g}usa,
North Cyprus, Mersin 10 - Turkey.}

\begin{abstract}
Spherically symmetric classical model of an elementary particle or a black
hole spacetime without central singularity had been constructed by O. B.
Zaslavskii in PRD 70(2004)104017. In this model an extremal Reissner-Nordstr%
\"{o}m (RN) black hole and a Bertotti-Robinson (BR) spacetime are glued at
the horizon such that the inner/core spacetime is the regular BR while
outside is the extremal RN. In this note we investigate the stability of
such a particle / regular black hole against linear radial perturbations.
The model turns out to be stable against such perturbations with a linear
equation of state after the perturbation.
\end{abstract}

\pacs{}
\maketitle

\section{Introduction}

With the advent of general relativity, the geometric theory of gravitation
due to Einstein, attempts to establish geometric model of an elementary
particle took start. Upon discovering a model spacetime in which the radius
of spacetime had a minimum encouraged Einstein and Rosen to propose such a
model of elementary particle \cite{ER}. Later on the model turned out to be
interpreted as an Einstein-Rosen bridge, and in present day's terminology as
a wormhole connecting two asymptotically flat spacetimes. Initially what
Einstein and Rosen called the radius of the elementary particle coincides
with the throat of the wormhole \cite{MT,MV}. The two spacetimes that are
connected at the throat may host black holes which are thought recently to
be entangled from the view of quantum interpretation \cite{Maldacena}.
Existence of singularities at the center of black holes has been a serious
obstacle in the construction of a self-consistent geometric model of a
particle. From physical grounds we observe / detect no singularity at the
location of a particle such as an electron. This automatically eliminates,
the models of black holes as viable particle models. Next, we must have a
smooth matching of the inner and outer geometries at the surface of the so
called particle model with well-defined energy scale that doesn't contradict
experiments. To overcome these requirements certain boundary conditions are
imposed which came to be known as the Israel's junction conditions \cite%
{Israel}. Such a construction was considered by Vilenkin and Fomin in \cite%
{Vilenkin1, Vilenkin2} and recently by Zaslavskii in \cite{Zaslavskii} where
he attempted to match different spacetimes to fit the physical requirements.
Naturally the choice of the regular Minkowski space as the inner core region
to external spacetimes such as Schwarzschild or Reissner-Nordstr\"{o}m (RN)
deserved a special attention. Such a choice, however, failed to satisfy the
matching conditions. The next attempt by Zaslavskii was to test another
regular candidate for the inner region for an elementary particle. The
choice has been the regular Bertotti-Robinson (BR) \cite{BR} spacetime that
represents a uniform electromagnetic field which is manifestly regular, to
be matched with the external RN geometry. The matching works consistently
provided the surface of the particle coincides with the event horizon of the
extremal RN geometry. As shown in \cite{Zaslavskii} the proper mass of the
particle turns out to be negative which vanishes identically with the
vanishing radius of the particle.

Our aim in this study is to investigate the stability of such a geometrical
model of an elementary particle. We perturb the junction shell which sits at
the horizon of the extremal RN black hole with BR spacetime as its inner
partner. Following the small radial perturbation a highly non-linear
equation is obtained which is plotted with chosen initial conditions. The
plots indicate that for tuned parameters the particle exposes stable
behaviours whereas its energy density remains negative. To develop analogues
models we investigate also the model with inner metrics other than BR. For
instance the choice of de Sitter serves our purpose well. The outer metric
is RN, which is more general than the extremal RN, and matches well with the
inner de Sitter spacetime. We have proposed this model recently as a method
to cut / eliminate the time like singularity of the RN geometry \cite{MH}.
It remains to be seen whether such an approach of cut and paste of
spacetimes on a thin-shell removes the singularities of other black holes.
It should be added also that in such a mechanism the physical parameters of
the external RN geometry are derived from geometrical properties, which is
reminiscent of the Wheeler's geometrodynamics \cite{Wheeler}.

\section{Stability of the Zaslavskii's particle model}

The geometrical particle model introduced by Zaslavskii in \cite{Zaslavskii}
consists of the core and outer spacetimes given by%
\begin{equation}
ds_{core}^{2}=-e^{2r/r_{0}}dt^{2}+dr^{2}+r_{0}^{2}\left( d\theta ^{2}+\sin
^{2}\theta d\phi ^{2}\right)
\end{equation}%
and%
\begin{multline}
ds_{outer}^{2}=-\left( 1-\frac{r_{h}}{r}\right) ^{2}dt^{2}+ \\
\frac{dr^{2}}{\left( 1-\frac{r_{h}}{r}\right) ^{2}}+r^{2}\left( d\theta
^{2}+\sin ^{2}\theta d\phi ^{2}\right)
\end{multline}%
respectively. These two regions are glued at $r=r_{0}$ on a timelike
hypersurface defined by%
\begin{equation}
F:=r-r_{0}\left( \tau \right) =0
\end{equation}%
where $\tau $ stands for the proper time, as described below.
Straightforward calculation gives the induced metric on the shell due to $%
ds_{core}^{2}$ as%
\begin{equation}
ds_{F}^{2}=-e^{2}dt^{2}+r_{0}^{2}\left( d\theta ^{2}+\sin ^{2}\theta d\phi
^{2}\right)
\end{equation}%
and $ds_{outer}^{2}$ 
\begin{equation}
ds_{F}^{2}=-\left( 1-\frac{r_{h}}{r_{0}}\right) ^{2}dt^{2}+r_{0}^{2}\left(
d\theta ^{2}+\sin ^{2}\theta d\phi ^{2}\right)
\end{equation}%
respectively. These clearly match provided the angular coordinates of both
spacetimes are identified on the shell and the coordinate time satisfies the
relations%
\begin{multline}
-e^{2}dt_{core}^{2}+dr^{2}= \\
-\left( 1-\frac{r_{h}}{r_{0}}\right) ^{2}dt_{outer}^{2}+dr^{2}=-d\tau ^{2}
\end{multline}%
with proper time $\tau $ on the shell. Having (6), the first fundamental
form of the shell is continuous across the shell. To see the situation of
the second fundamental form we start from the definition of the extrinsic
curvature%
\begin{equation}
K_{ij}=-n_{\mu }\left( \frac{\partial ^{2}x^{\mu }}{\partial \xi
^{i}\partial \xi ^{j}}+\Gamma _{\alpha \beta }^{\mu }\frac{\partial
x^{\alpha }}{\partial \xi ^{i}}\frac{\partial x^{\beta }}{\partial \xi ^{j}}%
\right)
\end{equation}%
in which 
\begin{equation}
n_{\mu }=\frac{1}{\sqrt{g^{\alpha \beta }\frac{\partial F}{\partial
x^{\alpha }}\frac{\partial F}{\partial x^{\beta }}}}\frac{\partial F}{%
\partial x^{\mu }}
\end{equation}%
is the normal $4-$vector which points from the shell outward for the
outer-side of the shell and inward for the inner-side of the shell. Explicit
calculation yields 
\begin{equation}
K_{icore}^{j}=diag\left( \frac{1+\dot{r}_{0}^{2}+r_{0}\ddot{r}_{0}}{r_{0}%
\sqrt{1+\dot{r}_{0}^{2}}},0,0\right)
\end{equation}%
while%
\begin{multline}
K_{iouter}^{j}= \\
diag\left( \frac{\left( 1-\frac{r_{h}}{r_{0}}\right) \frac{r_{h}}{r_{0}^{2}}+%
\ddot{r}_{0}}{\sqrt{\left( 1-\frac{r_{h}}{r_{0}}\right) ^{2}+\dot{r}_{0}^{2}}%
},\frac{\sqrt{\left( 1-\frac{r_{h}}{r_{0}}\right) ^{2}+\dot{r}_{0}^{2}}}{%
r_{0}},0\right) .
\end{multline}%
%
%
%
%
%
%
%
%
%
%
%
%
%
\begin{figure}[h]
\includegraphics[width=70mm,scale=0.7]{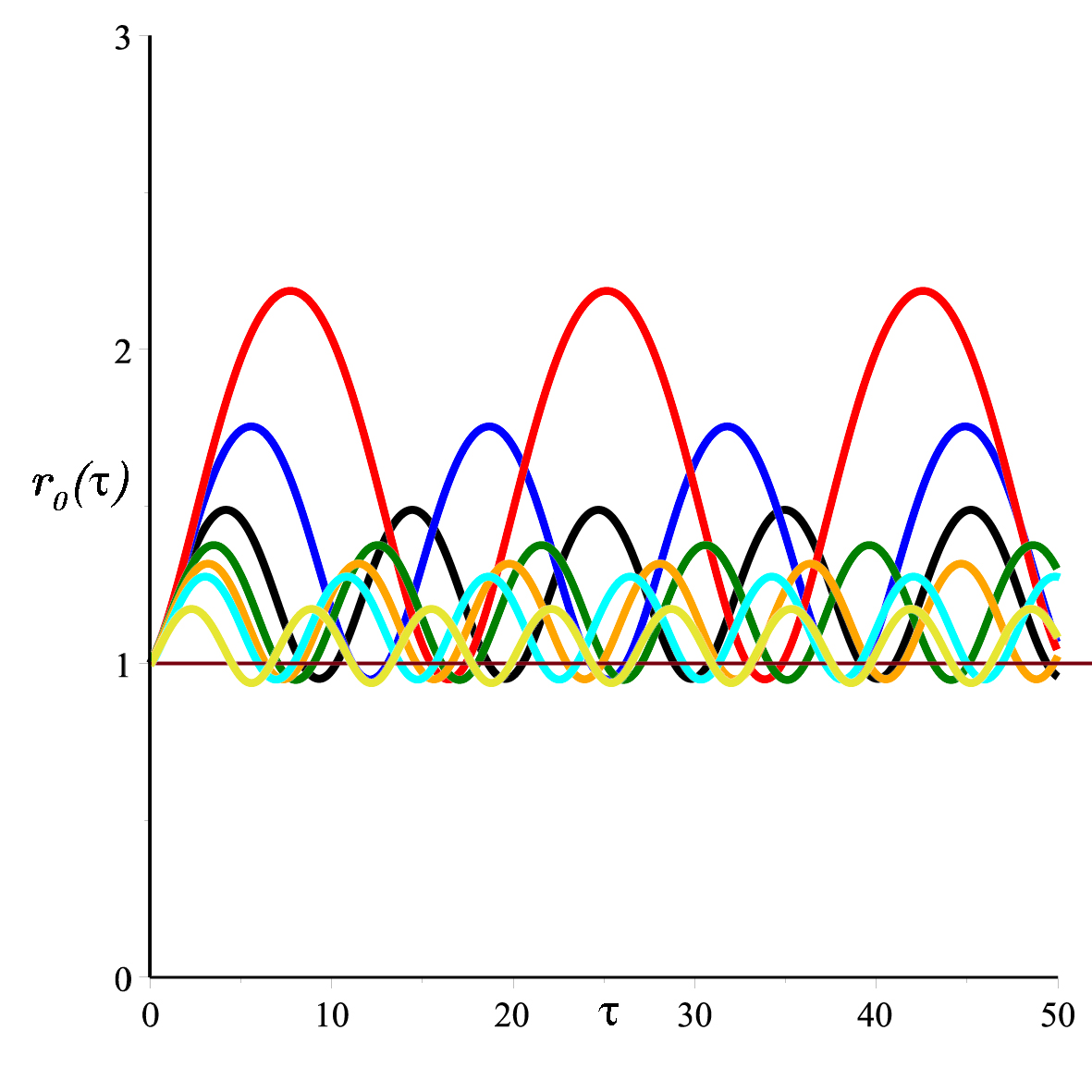}
\caption{A plot of $r_{0}\left( \protect\tau \right) $ versus $\protect\tau $
satisfying the equation of motion (16) with initial condition $r_{0}\left(
0\right) =1,\dot{r}_{0}\left( 0\right) =0.1.$ Different plots are for
different $\protect\omega =0.1,0.2,0.4,0.6,0.8,1.0$ and $2$. Here the bigger
the $\protect\omega $ the bigger the frequency. Note that $r_{h}=1$ is shown
as a horizontal line$.$ }
\end{figure}

\begin{figure}[h]
\includegraphics[width=70mm,scale=0.7]{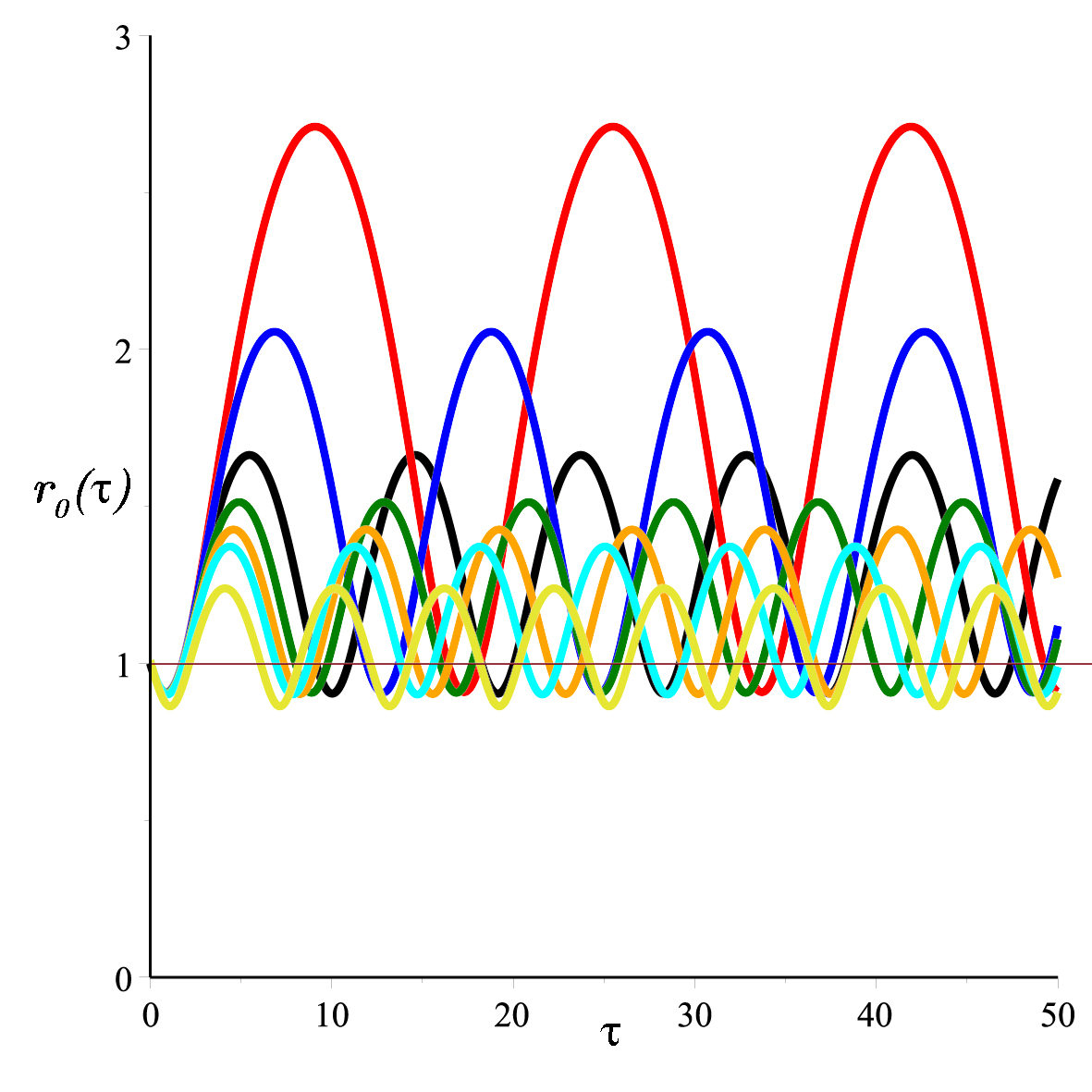}
\caption{A similar graph of $r_{0}\left( \protect\tau \right) $ versus $%
\protect\tau $ satisfying the equation of motion (16) with initial condition 
$r_{0}\left( 0\right) =1,\dot{r}_{0}\left( 0\right) =-0.2.$ Different plots
are for different $\protect\omega =0.1,0.2,0.4,0.6,0.8,1.0$ and $2$. Here
the bigger the $\protect\omega $ the bigger the frequency. Note that $r_{h}=1
$ is shown as a horizontal line$.$ }
\end{figure}
Applying the Israel junction condition \cite{Israel} 
\begin{equation}
-8\pi GS_{i}^{j}=\left[ K_{i}^{j}\right] -\left[ K\right] \delta _{i}^{j}
\end{equation}%
in which the bracket implies $\left[ X\right] =X_{outer}-X_{core},$ $%
K=trK_{i}^{j}$ and $S_{i}^{j}=diag\left( -\sigma ,p,p\right) $ we obtain ($%
8\pi G=1$)%
\begin{equation}
\sigma =-\frac{2\sqrt{\left( 1-\frac{r_{h}}{r_{0}}\right) ^{2}+\dot{r}%
_{0}^{2}}}{r_{0}}
\end{equation}%
and%
\begin{multline}
p=\frac{\left( 1-\frac{r_{h}}{r_{0}}\right) \frac{r_{h}}{r_{0}^{2}}+\ddot{r}%
_{0}}{\sqrt{\left( 1-\frac{r_{h}}{r_{0}}\right) ^{2}+\dot{r}_{0}^{2}}}+ \\
\frac{\sqrt{\left( 1-\frac{r_{h}}{r_{0}}\right) ^{2}+\dot{r}_{0}^{2}}}{r_{0}}%
-\frac{1+\dot{r}_{0}^{2}+r_{0}\ddot{r}_{0}}{r_{0}\sqrt{1+\dot{r}_{0}^{2}}}.
\end{multline}%
Let's note that at the static equilibrium condition where $r_{0}=const.\geq
r_{h}$ we get%
\begin{equation}
p_{0}=0
\end{equation}%
and 
\begin{equation}
\sigma _{0}=-\frac{2\left( 1-\frac{r_{h}}{r_{0}}\right) }{r_{0}}
\end{equation}%
which vanishes at $r_{0}=r_{h}$ \cite{Zaslavskii}.

Hence we assume that the system is at equilibrium at $r_{0}=r_{h}.$ Any
radial perturbation causes $r_{0}\neq r_{h}$ and therefore $p$ and $\sigma $
are obtained by (12) and (13) which means that on the perturbed shell there
would be a perfect fluid presented. Considering a linear equation of state
of the form $p=\omega \sigma $ ($\omega =const.$) one finds an equation of
motion given by%
\begin{multline}
\frac{\left( 1-\frac{r_{h}}{r_{0}}\right) \frac{r_{h}}{r_{0}^{2}}+\ddot{r}%
_{0}}{\sqrt{\left( 1-\frac{r_{h}}{r_{0}}\right) ^{2}+\dot{r}_{0}^{2}}}+ \\
\frac{\left( 1+\omega \right) \sqrt{\left( 1-\frac{r_{h}}{r_{0}}\right) ^{2}+%
\dot{r}_{0}^{2}}}{r_{0}}-\frac{1+\dot{r}_{0}^{2}+r_{0}\ddot{r}_{0}}{r_{0}%
\sqrt{1+\dot{r}_{0}^{2}}}=0
\end{multline}%
which is a differential equation giving the behaviour of $r_{0}$ after the
perturbation. This equation is not solvable analytically but one may purse a
numerical solution provided the initial conditions are known. As we consider
a radial perturbation about $r_{0}=r_{h}$ we set the following to be the
initial conditions: at $\tau =0,$ $r_{0}=r_{h}=1$ and $\dot{r}_{0}\neq 0.$
In Figs. 1 and 2 we plot the solution of Eq. (16) for $\dot{r}_{0}=0.1$ and $%
-0.2,$ respectively, and $\omega =0.1,0.2,0.4,0.6,0.8,1.0$ and $2.0.$ As we
observe here; for a positive $\omega $ and a small perturbation the particle
remains stable. In both cases, the shell spends more time with $r_{0}>r_{h}$
although in Fig. 2 initially the shell is perturbed toward the center. In
such a particle model, however, both the energy density $\sigma $ and
angular pressure $p$ are negative. They both vanish at the equilibrium
radius while perturbation drives the system to attain $\sigma <0$ and $p<0.$
Assuming that the parameter $\omega <0,$ yields a model with $\sigma <0$ and 
$p>0$ which results unfortunately in an unstable particle model. At any cost
Eq. (12) implies that the model will always have $\sigma <0.$

From (16) at $r_{0}=r_{h}=1$ one finds the acceleration%
\begin{equation}
a=\left( 1-\frac{\omega \left\vert v\right\vert }{\sqrt{1+v^{2}}-\left\vert
v\right\vert }\right) \left\vert v\right\vert \sqrt{1+v^{2}}
\end{equation}%
in which $v=$ $\dot{r}_{0}$ and $a=\ddot{r}_{0}$ both at $r_{0}=r_{h}=1.$
The direction of acceleration $a$ depends not only on the magnitude of $v$
but also on the value of $\omega .$ Hence for the case that directions of $v$
and $a$ are both negative the shell is more likely to become unstable. For
the negative $\omega $, with positive initial velocity the shell is also
unstable and that is why we chose $\omega >0$ in Figs. 1 and 2.

\section{A general overview of the particle model}

%
%
%
%
%
%
%
\begin{figure}[h]
\includegraphics[width=70mm,scale=0.7]{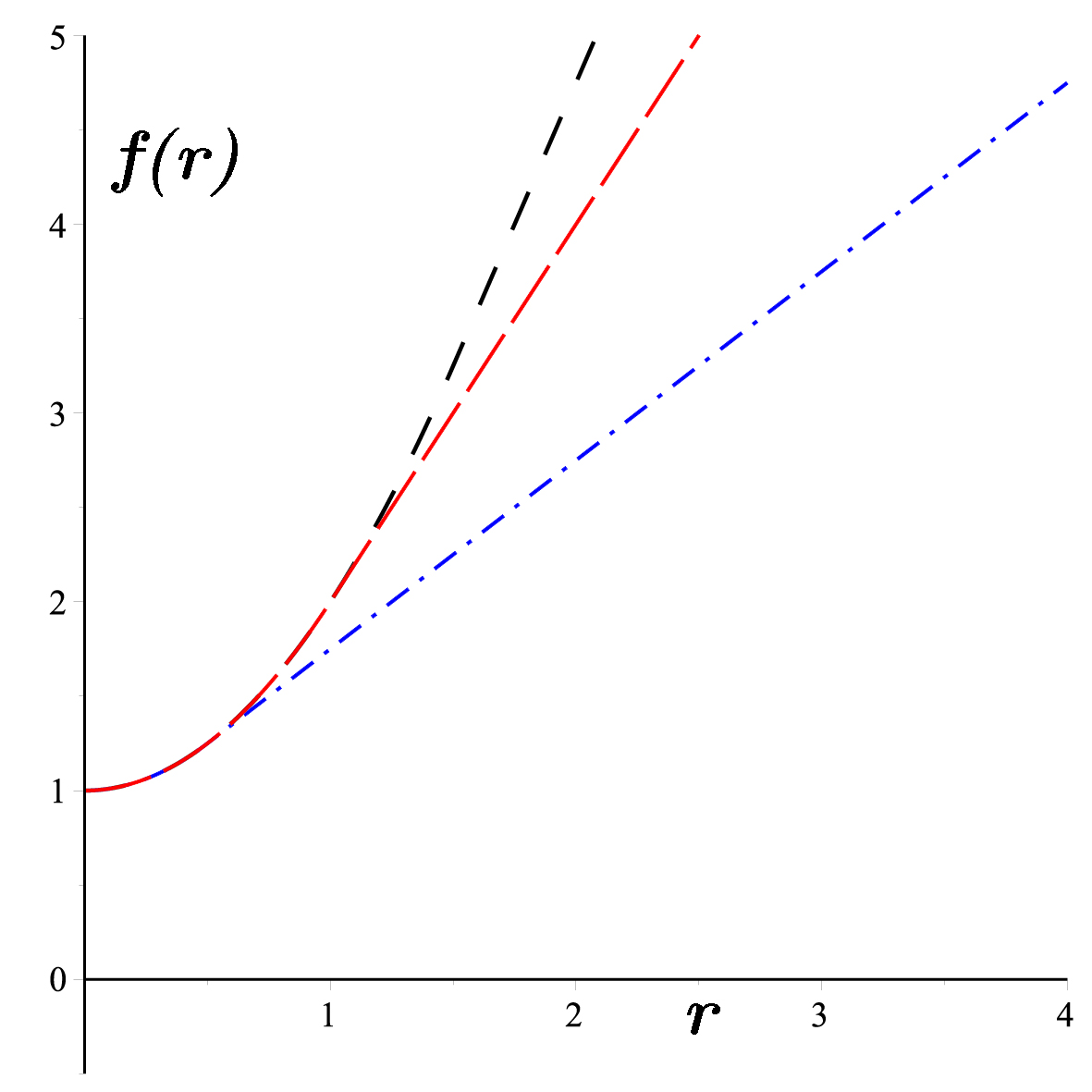}
\caption{Metric function $f\left( r\right) $ vs $r$ in accordance with Eq.
(30) and (31) for $\ell ^{2}=1$ and $R_{0}=0.5,1.0$ and $1.5.$ }
\end{figure}
%
%
%
%
%
%
%

To complete our study we consider the static spherically symmetric
spacetimes in general form. The line element inside and outside the shell of
the particle are given by%
\begin{equation}
ds^{2}=-f_{i}\left( r_{i}\right) dt_{i}^{2}+\frac{dr_{i}^{2}}{f_{i}\left(
r_{i}\right) }+r_{i}^{2}\left( d\theta _{i}^{2}+\sin ^{2}\theta _{i}d\phi
_{i}^{2}\right)
\end{equation}%
with $i=1,2$ indicating the inside and outside, respectively. The energy
momentum tensor components on the shell are found to be from $S_{\mu }^{\nu
}=diag\left( -\sigma ,p,p\right) $%
\begin{equation}
\sigma =-\frac{1}{4\pi G}\left( \frac{\sqrt{f_{2}\left( R\right) +\dot{R}^{2}%
}-\sqrt{f_{1}\left( R\right) +\dot{R}^{2}}}{R\left( \tau \right) }\right)
\end{equation}%
and%
\begin{multline}
p=\frac{1}{8\pi G}\left( \frac{2\ddot{R}\left( \tau \right) +f_{2}^{\prime
}\left( R\right) }{2\sqrt{f_{2}\left( R\right) +\dot{R}^{2}}}-\frac{2\ddot{R}%
\left( \tau \right) +f_{1}^{\prime }\left( R\right) }{2\sqrt{f_{1}\left(
R\right) +\dot{R}^{2}}}\right. + \\
\left. \frac{\sqrt{f_{2}\left( R\right) +\dot{R}^{2}}-\sqrt{f_{1}\left(
R\right) +\dot{R}^{2}}}{R\left( \tau \right) }\right)
\end{multline}%
in which $r=R\left( \tau \right) $ represents the surface of the particle.
In static condition one finds%
\begin{equation}
\sigma _{0}=-\frac{1}{4\pi G}\left( \frac{\sqrt{f_{2}\left( R_{0}\right) }-%
\sqrt{f_{1}\left( R_{0}\right) }}{R_{0}}\right) ,
\end{equation}%
and%
\begin{multline}
p_{0}=\frac{1}{8\pi G}\left( \frac{f_{2}^{\prime }\left( R_{0}\right) }{2%
\sqrt{f_{2}\left( R_{0}\right) }}-\frac{f_{1}^{\prime }\left( R_{0}\right) }{%
2\sqrt{f_{1}\left( R_{0}\right) }}\right. + \\
\left. \frac{\sqrt{f_{2}\left( R_{0}\right) }-\sqrt{f_{1}\left( R_{0}\right) 
}}{R_{0}}\right) .
\end{multline}%
Considering $\sigma _{0}=0$ imposes 
\begin{equation}
f_{2}\left( R_{0}\right) =f_{1}\left( R_{0}\right)
\end{equation}%
while $p_{0}=0$ yields%
\begin{equation}
f_{2}^{\prime }\left( R_{0}\right) =f_{1}^{\prime }\left( R_{0}\right) .
\end{equation}%
Having these conditions satisfied, the particle model becomes physical. In 
\cite{MH} we have shown that the outer RN black hole can be glued to the
inner de-Sitter spacetime consistently and due to that we proposed to remove
the singularity of the RN black hole.

For instance, if we set $f_{1}\left( r\right) =1-\frac{r^{2}}{\ell ^{2}}$
and $f_{2}\left( r\right) =\left( 1-\frac{e}{r}\right) ^{2}$ the latter
equations become%
\begin{equation}
1-\frac{R_{0}^{2}}{\ell ^{2}}=\left( 1-\frac{e}{R_{0}}\right) ^{2}
\end{equation}%
and%
\begin{equation}
-\frac{R_{0}}{\ell ^{2}}=\frac{e}{R_{0}^{2}}\left( 1-\frac{e}{R_{0}}\right)
\end{equation}%
respectively. Simultaneous solution of these equations reveals that 
\begin{equation}
R_{0}=\frac{2}{3}e
\end{equation}%
which is smaller than the horizon radius i.e., $r_{h}=e$. In addition to
that one obtains the geometrical condition 
\begin{equation}
\frac{R_{0}^{2}}{\ell ^{2}}=\frac{3}{4}.
\end{equation}

One simple solution for these equations can be found by considering Taylor
expansion of the outer metric about $R_{0}$ i.e., 
\begin{equation}
f_{2}\left( r\right) \simeq f_{2}\left( R_{0}\right) +f^{\prime }\left(
R_{0}\right) \left( r-R_{0}\right) .
\end{equation}%
In this case we set 
\begin{equation}
f_{1}\left( r\right) =1+\frac{r^{2}}{\ell ^{2}}
\end{equation}%
\ and due to (24) and (25) we find $f_{2}\left( R_{0}\right) =f_{1}\left(
R_{0}\right) =1+\frac{R_{0}^{2}}{\ell ^{2}}$ and $f_{2}^{\prime }\left(
R_{0}\right) =f_{1}^{\prime }\left( R_{0}\right) =\frac{2R_{0}}{\ell ^{2}}$
so that to the first order we find%
\begin{equation}
f_{2}\left( r\right) =1-\frac{R_{0}^{2}}{\ell ^{2}}+\frac{2R_{0}}{\ell ^{2}}%
r.
\end{equation}%
The spacetime expressed by the latter metric is singular \cite{GM} at $r=0$.
This singularity may be naked with $\ell ^{2}\geq R_{0}^{2}$ or hidden
behind a horizon for $\ell ^{2}<R_{0}^{2}$ located at 
\begin{equation}
r_{h}=\frac{R_{0}^{2}-\ell ^{2}}{2R_{0}}.
\end{equation}%
As one expects for the particle model $R_{0}\geq r_{h}$ and one finds%
\begin{equation}
R_{0}^{2}\geq -\ell ^{2}
\end{equation}%
which is trivially satisfied. Let's add that the energy momentum tensor of
the spacetime outside the shell is found to be of the perfect fluid form
given by%
\begin{equation}
T_{\mu }^{\nu }=diag\left( -\rho ,P_{r},P_{\theta },P_{\phi }\right) 
\end{equation}%
in which%
\begin{equation}
\rho =-\frac{4R_{0}}{r\ell ^{2}}+\frac{R_{0}^{2}}{r^{2}\ell ^{2}},
\end{equation}%
\begin{equation}
P_{r}=-\rho 
\end{equation}%
and%
\begin{equation}
P_{\theta }=P_{\phi }=\frac{2R_{0}}{r\ell ^{2}}.
\end{equation}%
In Fig. 3 we plot $f\left( r\right) $ versus $r$ for $\ell ^{2}=1$ and $%
R_{0}=0.5,1.0$ and $1.5$ from bottom to the top. The smooth matching is
clearly seen from the plots.

\section{Conclusion}

We revisit the geometrical model of an elementary particle \cite{Zaslavskii}
with the purpose to investigate its stability and search for alternative
models. The suggested model in \cite{Zaslavskii} glued spacetimes BR
(inside) with the extremal RN (outside). The radius of the elementary
particle coincides with the horizon of the extremal RN. At the static
equilibrium the shell has no surface energy-momentum, i.e., $\sigma
_{0}=p_{0}=0.$ Upon radial perturbation the shell absorbs energy and behaves
as a perfect fluid satisfying $p=\omega \sigma .$ When the shell is
perturbed a master equation (i.e., Eq. (16) for the general spherical
symmetry configuration, given in Sec. III), governing the perturbations is
derived and plotted for different initial conditions. The oscillatory
behaviours in Figs. 1 and 2 support the stability of the shell and therefore
the particle is stable against such perturbations. An inward motion is
observed to reverse immediately outward following a small penetration of the
shell below the horizon. The energy density attained by the shell in this
process is negative. The particle model constructed from BR and RN can be
modified by invoking other regular metrics to replace the BR. One immediate
choice is the de Sitter metric to be matched with the outer RN metric. In
this venture naturally the mass and charge of the RN metric are defined
entirely from the geometrical parameters.

\end{document}